\documentclass[a4paper, 11pt]{article} 

\usepackage{mathpazo} 
\usepackage{longtable}
\usepackage{hyperref}
\usepackage[numbers,sort&compress]{natbib}

\makeatletter

\renewcommand{\maketitle}{ 
\begin{flushright} 
{\LARGE\@title} 

\vspace{50pt} 

{\large\@author} 
\\\@date 

\vspace{40pt} 
\end{flushright}
}

\title{\textbf{Reconsidering Written Language}}

\author{\textsc{Gopal Sarma} 
\\{\textit{gopalsarma@outlook.com}}} 

\date{} 

\begin{document}

\maketitle 



\begin{abstract}
A number of elite thinkers in Europe during the 16th and 17th centuries pursued an agenda which historian Paolo Rossi calls the ``quest for a universal language,'' a quest which was deeply interwoven with the emergence of the scientific method.  From a modern perspective, one of the many surprising aspects of these efforts is that they relied on a diverse array of memorization techniques as foundational elements.  In the case of Leibniz's \emph{universal calculus}, the ultimate vision was to create a pictorial language that could be learned by anyone in a matter of weeks and which would contain within it a symbolic representation of all domains of contemporary thought, ranging from the natural sciences, to theology, to law.  In this brief article, I explore why this agenda might have been appealing to thinkers of this era by examining ancient and modern memory feats.  As a thought experiment, I suggest that a society built entirely upon memorization might be less limited than we might otherwise imagine, and furthermore, that cultural norms discouraging the use of written language might have had implications for the development of scientific methodology.  Viewed in this light, the efforts of Leibniz and others seem significantly less surprising.  I close with some general observations about cross-cultural origins of scientific thought.  
\end{abstract}


\vspace{15pt} 

 
\section*{Introduction}
The growth of the scientific method in 16th and 17th century Europe took place amidst an unusual cultural milieu in which schools of thought on memory and memorization were at the forefront of intellectual developments.  During this time period, memory was viewed not only as a set of practices for remembering, but also as a foundational methodology for structuring knowledge \cite{yates, rossi, sarma}. \\

Whereas in modern times memorization is often thought of as being antithetical to conceptual thinking, Renaissance and Englightenment era European culture gave rise to significant innovation in the techniques and applications of memorization.  Building on ideas originating in ancient Greece and Rome, European intellectuals from several centuries ago developed highly sophisticated visualization techniques that allowed them to recall massive amounts of information with astonishing accuracy.  And in the minds of some of the greatest scientific and philosophical innovators of the time period, notably Francis Bacon, Renes Descartes, and Gottfriend Leibniz, as well as other lesser known thinkers, these memorization techniques formed the basic set of concepts around which many thoughts about scientific methodology were formed.  \\

The core premise of this argument- that memory and memorization played a critical role in the emergence of scientific culture- is that the development of the scientific method was not a discrete event. Principled reasoning and systematic investigation have always been part of human society, but during this time period, they became distilled and adopted at an institutional level. How did this transformation take place and why had it not before? In a sense, what forms scientific reasoning consists of a complex and rich set of cultural norms, but which at its core, is rather mundane. By nature, people procrastinate, cut corners, and are not systematic in their efforts. Scientific reasoning might be described as a collection of systematic efforts conducted in a context that prioritizes an attempt to uncover the basic principles governing the behavior of the natural world. How then did the adoption of a systematic approach to knowledge accelerate during the time period traditionally associated with the scientific revolution? \\

In a previous article, I argued that the art of memory provided an inspiring vision of what could be accomplished with a systematic approach to knowledge \cite{sarma}. While the basic notions of scientific reasoning may not be abstract, they are not necessarily easy to put into practice, and furthermore, it is not always clear what precise steps would even constitute reasonable next actions for an institution attempting to transform itself into a more rigorous, long-term research establishment.  In the absence of a critical mass of scientific accomplishments, major intellectual shifts would have also been difficult to justify. The notion of a journal system and peer review seem elementary to us, but these ideas are far from obvious and required real momentum to become part of the bedrock of institutional practice. At this critical juncture in human history, the art of memory provided a clear, tangible vision and concrete motivation for widespread adoption of a more systematic approach to knowledge.  On the relationship of Francis Bacon and Renes Descartes' thinking to the art of memory, for instance, Paolo Rossi writes:
\begin{quote}
{\small \noindent In the works of Descartes and Bacon there is evidence of a direct knowledge of sixteenth-century works on mnemonics.  Bacon often mentions 'collection of loci', 'syntaxes' and 'artificial memory', and makes explicit references to the 'doctrine of loci', the 'collocation of images' and the Lullian 'typocosmia'.  Descartes, who makes far fewer explicit references and generally refrains from quotation, nonetheless mentions having read Schenkel's Ars memorativa, and often refers to the ars memoriae and the role played by 'sensible images' in the representation of intellectual concepts.  He also refers to the Lullian idea of the 'chain of the sciences' (catena scientiarum), mentions his particular interest in one of Lull's anonymous followers, and asked his friend Isaac Beeckmann about the publication of Agrippa's Lullist works, and discussed with him the significance and potential of the Lullian art.  The themes of the ars memorativa and ars combinatoria seem to have exercised a significant influence on the thought of Bacon and the young Descartes, and some of these themes can be related directly to Bacon's conception of a 'new logic' and Descartes's 'new method' of philosophizing \cite[p.103]{rossi}.}
\end{quote}

And yet, despite the intense interest of these visionary thinkers, the most ambitious paradigms of this time period did not materialize.  It is difficult to imagine that Camillo's memory theater- an effort to create systematized, physical figurines representing a taxonomy of all knowledge- Bruno's combinatory wheels, or Leibniz' universal calculus would have delivered on their promises. But in spite of their failures, these efforts shared the common feature of representing a systematic approach to knowledge.  The art of memory and efforts to develop a universal language provided a vision that helped leading intellectuals to clarify and frame their thoughts and to catalyze the adoption of scientific reasoning on a large scale.  It created a philosophical context which allowed the more standard and widely discussed factors- patronage, the rise of a journal system, an explicitly articulated notion of hypothesis-driven investigation, etc.- to take root at an institutional level.

\section*{Memorization and Written Language}
In modern times, there is a standard refrain about the role of memory and memorization in previous eras, namely that in the absence of convenient methods for writing, a trained memory would have been practically significant in ways that it is not today.  But something noteworthy about the European schools of thought on memory is that they continued to exist several centuries after the development of the printing press.  In examining the culture of this time period, it is clear that memory was viewed in a very different light than by thinkers of our own era.  Most broadly, memory was thought of by many 16th and 17th century intellectuals as being a foundational methodology for structuring knowledge.  Furthermore, a few key individuals, notably Gottfried Leibniz, Giordano Bruno, and Renes Descates, conceived of memory as being intimately related to developing symbolic means for representing scientific concepts.  For Leibniz, the mnemonic method was a critical element of his vision to develop a universal calculus, a symbolic language to represent and eliminate logical contradictions from the entirety of human knowledge.  \\

While the ambitions of creating a symbolic representation of knowledge is understandable from a modern perspective- certainly, mathematics itself is a symbolic language and has proven itself to be valuable in a wide variety of subjects- what is perplexing about this agenda is the emphasis on memorization.  Why was this perspective particularly appealing to thinkers of this time period, when in contemporary society, many practical technological tools- from address books, to electronic calendars, to smart phones- are specifically aimed at relieving our minds from the burden of memory? 
With the aim of elucidating the mindset of the time-period, I will suggest the following counterfactual thought experiment.  What would a world look like in which efforts analogous to those of Leibniz and his contemporaries thoroughly penetrated all sectors of society?  What would be the consequences for a society that relied on memorization for both daily and intellectual tasks, rather than written language?  \\

To imagine this possibility, we can simply return back in time to several thousand years ago before writing systems were widely adopted.  In particular, it seems as though, around the time when written language came into usage, strong cultural norms \emph{discouraging its use} might have precipitated analogous developments to those that took place in 16th and 17th century Europe.  Whereas several millennia had passed between the earliest use of mnemonic techniques in ancient Greece and their flowering in Renaissance and post-Renaissance Europe, a society which resisted the adoption of written language might have seen a comparable developmental trajectory- out of necessity- compressed into a much shorter time period.  The result might have been an intellectual explosion, a scientific revolution that is, but of a very different kind than what ultimately came to pass in Europe several millennia later.  \\

Rather than being merely a counter-factual historical curiosity, there is reason to believe that such a set of cultural norms might have actually emerged.  The critical observation is that written language- unlike its spoken counterpart- was an invention, and like with all inventions, would have been met with some amount of resistance.  Some of the animosity towards the written word came from highly informed, intellectual, and influential leaders, and were this resistance to have achieved a critical mass, the adoption of written language might have slowed considerably.  The following beautiful passage, from Socrates' \emph{Phaedrus}, illustrates this point:
\begin{quote}
{\small
I heard, then, that at Naucratis, in Egypt, was one of the ancient gods of that country, the one whose sacred bird is called the ibis, and the name of the god himself was Theuth. He it was who invented numbers and arithmetic and geometry and astronomy, also draughts and dice, and, most important of all, letters. Now the king of all Egypt at that time was the god Thamus, who lived in a great city of the upper region, which the Greeks call the Egyptian Thebes, and they call the god himself Ammon. To him came Theuth to show his inventions, saying that they ought to be imparted to the other Egyptians. But Thamus asked what use there was in each, and as Theuth enumerated their uses, expressed praise or blame of the various arts which it would take too long to repeat; but when they came to letters, 'This invention, O king,' said Theuth, 'will make the Egyptians wiser and will improve their memories; for it is an elixir of memory and wisdom that I have discovered.' But Thamus replied, 'Most ingenious Theuth, one man has the ability to beget arts, but the ability to judge of their usefulness or harmfulness to their users belongs to another; and now you, who are the father of letters, have been led by your affection to ascribe to them a power the opposite of that which they really possess. For this invention will produce forgetfulness in the minds of those who learn to use it, because they will not practise their memory. Their trust in writing, produced by external characters which are not part of themselves will discourage the use of their own memory within them. You have invented an elixir not of memory but of reminding; and you offer your pupils the appearance of wisdom, not true wisdom, for they will read many things without instruction and will therefore seem to know many things, when they are for the most part ignorant and hard to get along with, since they are not wise, but only appear wise \cite{socrates}.}
\end{quote}

If skepticism of influential leaders such as King Thamus were more widespread and if opposition to written language had reached a critical mass, what might the resulting society have looked like?  We would have to imagine that there would be a division of labor entirely devoted to the maintenance of different types of knowledge.  Differences in memory that might appear to be extremely subtle to us would be brought to the forefront.  There might be groups of people responsible for maintaining long-term knowledge- say, related to agriculture and medicine, or literature even- and others responsible for knowledge that is overturned more quickly, for example, the inventory of vendors in a marketplace.  Furthermore, it seems quite likely that the need to maintain all knowledge in memory would have created a natural and organic selective pressure towards infrastructural simplicity, particularly in the construction of legal and political systems. \\

To further lend evidence for the plausibility of such a society, I have listed in Table 1 various memory feats that have been performed in ancient and in modern times.  The great epics from Greece and India, the Iliad and the Mahabharata, were largely carried down in an oral tradition and were unlikely to have been written down for a long period after they were composed.  These were almost certainly memorized using the standard techniques of repetition, although it is worth mentioning that the meter and poetic structure of these epics were significant aids to the memory as well.  In our own era, the art of memory has largely been confined to the domain of a rather peculiar set of ``memory competitions''\footnote{Joshua Foer's memoir Moonwalking with Einstein is a beautiful first person account of the history, techniques, and personalities of the ``competitive memory circuit'' \cite{foer}.  Perhaps one of the primary lessons of Foer's book relevant to this article is that most of the participants in the various international memory competitions do not claim to have strong natural memories.  Rather, their memories were highly trained in the specific context of tasks relevant to competitive memory, for example, memorizing a long list of historical dates, or memorizing the order of a deck of cards.  This observation lends further plausibility to the idea of an entire society in which such techniques were commonplace and in which written language was discouraged- these techniques would easily have been learned by many, and need not have been restricted to a handful of elites.}  in which participants are challenged with a broad array of memorization tasks that seem almost superhuman to those unfamiliar with mnemonic techniques.  In the context of the present article, I mention these competitions to give an example of the highly diverse kinds of information that practitioners have developed specialized techniques to remember.  For a society built upon memorization, techniques such as these would have no doubt been commonplace- as they were during the Middle Ages, the Renaissance, and the Enlightenment- and one might imagine that for the sake of redundancy and error correction, there would have been social norms encouraging different groups to use different techniques to maintain a given body of information.  The ancient art of memory and its recent reincarnation in the form of the competitive memory circuit suggest that both long-term, slowly evolving knowledge, as well as information that is rapidly overturned might have been recorded, processed, re-evaluated, and disseminated without the use of written language.

\begin{center}
\begin{longtable}{c | c | c}
\caption{Ancient and modern memory feats \cite{homer, mahabharata, goddard, foer}} \\
Task & Size of text / time & Techniques utilized \\
\hline
Iliad & $\approx$ 15,000 lines & Repetition / poetic structure \\
Mahabharata & $\approx$ 200,000 lines & Repetition / poetic structure \\
Speeches of Cicero\footnote{For historical reasons, Cicero's speeches have the distinction of having been memorized through a variety of different techniques \cite{yates, rossi, sarma}.} & Varied & Mnemonic method / repetition \\
One hour cards\footnote{In one hour cards, participants memorize the order of as many decks of cards as possible in one hour.} & 1456 (28 decks) & Mnemonic method \\
One hour numbers\footnote{In one hour numbers, participants memorize as many decimal digits as possible in one hour.} & 2660 digits & Mnemonic method \\
Speed cards (1 deck)\footnote{In speed cards, participants are given a single shuffled deck of cards (which has been duplicated in another deck) and the task is to memorize the order of all 52 cards in under 5 minutes.} & 21.19 seconds & Mnemonic method \\
Historic dates (5 min.)\footnote{In historic dates, participants are given a list of fictional events and years and asked to recall as many events as possible after a 5 min training period.} & 120 & Mnemonic method
\end{longtable}
\end{center}
It is also worth noting that the consistent usage of mnemonic techniques would have had a measurable impact on brain function and neural development as well.  It has been demonstrated, for example, that in contrast to memorization by ordinary repetition, use of the mnemonic method activates those regions of the brain otherwise responsible for spatial navigation \cite{maguire}.  Unsurprisingly, this is due to the fact that the mnemonic method relies explicitly on the use of spatially located images to form memories.  Indeed, it is interesting to note that during the Enlightenment, one would have been able to distinguish between disciples of the different schools of thought on memory simply via an fMRI!  That is, whereas the dialectic method of Petrus Ramus and his followers would not have given rise to a strong response in the visual regions of the brain, we would expect strong signals from parts of the visual cortex in those practitioners of the mnemonic method, as well as Bruno and Leibniz's hybrid methods.  For a society whose very operational foundation was built upon these techniques, one would expect systematic deviations in cognitive organization from an otherwise normal population.  One wonders if these cognitive differences in the capacity to generate intense imagery might have had derivative effects on creativity as well.  \\

While I have presented the above set of circumstances as a counter-factual historical thought experiment, it is worth considering that a scenario such as this one is what Leibniz anticipated would be the societal consequences of the universal calculus, as illustrated in the following excerpt:

\begin{quotation}
{\small
\noindent My invention contains all the functions of reason: it is a judge for controversies; an interpreter of notions; a scale for weighing probabilities; a compass which guides us through the ocean of experience; an inventory of things; a table of thoughts; a microscope for scrutinizing things close at hand; an innocent magic; a non-chimerical cabala; a writing which everyone can read in his own language; and finally a language which can be learnt in a few weeks, traveling swiftly across the world, carrying the true religion with it, wherever it goes \cite{leibniz}.
}
\end{quotation}

Aside from the colorful description of the potential impact of this ``invention,'' what is most striking is the astonishing time frame on which Leibniz believed this language could be acquired.  For a person to learn a foreign language in a few weeks would be a superhuman feat, and Leibniz was conceiving of a language that would be acquired on such a time scale which contained not only means for communicating ordinary concepts, but the entirety of human knowledge.  In other words, Leibniz was imagining a form of communication that was far superior to ordinary written language.  Ultimately, it was hoped that by taking inspiration from the geometric and pictorial aspects of the Chinese alphabet and Egyptian hieroglyphics, and fusing them with the memory-aiding power of the mnemonic method and related memorization techniques, a rigorous process of ``notation engineering'' would allow for the creation of a symbolic system that transparently represented the factual content as well as the logical structure of the entirety of human thought.  It hardly needs to be stated that this vision did not materialize, but it is a striking and original vision and one worth contemplating.  

\section*{Discussion}
If we accept the possibility of a society built upon memorization, and furthermore, that such a set of cultural norms might have had consequences for the development of scientific methodology and scientific culture, we would also have to accept the conclusion that a scientific revolution in such circumstances would have looked very different from what took place in Europe during the 16th and 17th centuries.  For instance, while we traditionally associate the scientific method with hypothesis-driven investigation, one of the primary innovations of the 17th century, and which subsequently formed the bedrock of the physical sciences, was Newton and Leibniz's infinitesimal calculus- on its own, a strictly mathematical theory that is only incidentally related to experimental science and hypothesis-driven investigation. If a scientific revolution were to have taken place at a very different place and time in human history, what might it have actually looked like?  What would have been the topics that received the most attention and what would the consequences have been?  Forcing ourselves to reason explicitly about such unusual circumstances might help to develop more precise models for what we mean by scientific reasoning, and help to disentangle larger principles from the specific set of historical circumstances in which those principles emerged. \\

Consider, for example, the recent interest in a scientific study of yoga, Tai Chi, and the meditative practices of the Tibetan Buddhists (see for example \cite{yoga, taichi, llama}).  Each of these ancient traditions is of sufficient intellectual depth to have drawn the attention of modern scientists.  And yet these practices came about from places and during time periods where there were no scientific journals, no means of rapid, widespread communication and double blind peer review, and without the infrastructure of the modern research university.  Clearly then, these early Indian, Chinese, and Tibetan thinkers and experimenters developed a highly sophisticated scientific culture in the absence of many common features of our own scientific world.  What would have happened if these efforts were to have been scaled up even more dramatically centuries and millennia ago?  What would have the resulting societies and intellectual culture looked like?  \\

In examining the astonishing developments taking place in the contemporary scientific world (see for example \cite{nature, nielsen}), there may be real value in encouraging the writing of counter-factual scientific histories (or at the very least thinking about them), and in particular, asking the question of what a scientific revolution might have looked like at different places and times in human history. The intuition gained from such exercises could prove useful in developing policy recommendations for and guiding the evolution of younger scientific institutions, particularly in the developing world, where there is likely to be more freedom and opportunity to influence the development of institutional culture. 

\section*{Acknowledgements}
I would like to thank Doug Bemis and Aaswath Raman for insightful discussions and critical reading of the manuscript.

\end{document}